\documentclass[12pt]{article}
\usepackage{graphicx}
\usepackage{lscape}
\usepackage{epsfig}
\usepackage{citesort}
\usepackage{amssymb}
\usepackage{amsmath}
\usepackage{multirow}
\usepackage[table]{xcolor}
\usepackage{colortbl}
\definecolor{lightgray}{gray}{0.9}

\newcommand{\be}{\begin{equation}}
\newcommand{\ee}{\end{equation}}
\newcommand{\bea}{\begin{eqnarray}}
\newcommand{\eea}{\end{eqnarray}}

\def\R1{\varepsilon_1}
\def\E8{\varepsilon_8}

\def\s1{\hat s}

\newcommand{\bd}{\begin{displaymath}}
\newcommand{\ed}{\end{displaymath}}

\newcommand{\f}{\frac}

\def\R1{\varepsilon_1}
\def\E8{\varepsilon_8}

\def\beq{\begin{equation}}
\def\eeq{\end{equation}}
\def\bea{\begin{eqnarray}}
\def\eea{\end{eqnarray}}
\def\beeq{\begin{eqnarray}}
\def\eeeq{\end{eqnarray}}

\def\nnb{\nonumber}

\def\rar{\rightarrow}
\def\nnb{\nonumber}

\def\ba{\begin{array}}
\def\ea{\end{array}}

\def\xis0{{\Xi^{*0}}}

\def\g5{\gamma_5}

\def\es{\!\!\! &=& \!\!\!}

\def\ar{&+& \!\!\!}
\def\ek{&-& \!\!\!}

\begin{document}
\title{
         {\Large
                 {\bf Analysis of the radiative $\Lambda_b
\rightarrow \Lambda \gamma$ transition
in SM and scenarios with one or two universal extra dimensions 
                 }
         }
      }
      
\author{\vspace{1cm}\\
{\small  K. Azizi$^1$ \thanks {e-mail: kazizi@dogus.edu.tr}\,\,, S. Kartal$^2$ \thanks
{e-mail: sehban@istanbul.edu.tr}\,\,, A. T. Olgun$^2$ \thanks
{e-mail: a.t.olgun@gmail.com}\,\,, Z. Tavuko\u glu$^2$ \thanks
{e-mail: z.tavukoglu@gmail.com}}  \\
{\small $^1$ Department of Physics, Do\u gu\c s University,
Ac{\i}badem-Kad{\i}k\"oy, 34722 \.{I}stanbul, Turkey}\\
{\small $^2$ Department of Physics, \.{I}stanbul University,
Vezneciler, 34134 \.{I}stanbul, Turkey}\\
}

\date{}
        \begin{titlepage}
\maketitle
\thispagestyle{empty}
\begin{abstract}
We investigate the radiative process of the $\Lambda_b
\rightarrow \Lambda \gamma$ in the standard model as well as models with one or two compact universal extra dimensions. Using the form factors entered to the low energy matrix elements, calculated via light cone QCD in full theory, we calculate the total decay width and branching ratio of this decay channel. We compare the results of the extra dimensional models with those of the standard model on the considered physical quantities and look for the deviations of the results from the standard model predictions at different values of the compactification scale ($1/R$). 
\end{abstract}

~~~PACS number(s): 12.60.-i, 13.30.-a, 13.30.Ce, 14.20.Mr 
\end{titlepage}


\section{Introduction}  
As it is well-known, the flavor changing neutral current (FCNC) transitions are prominent tools to indirectly search for the new physics (NP) effects. 
There are many mesonic and baryonic processes based on the $b \rightarrow s$ transition at quark level investigated in the literature via different NP 
models and compared the obtained results with the experimental data to put constraints on the NP parameters. One of the most important channels in agenda 
of different experimental groups is the baryonic FCNC $\Lambda_b \rightarrow \Lambda \ell^+\ell^-$ decay channel. The CDF Collaboration at Fermilab reported
 the first observation on this mode at muon channel \cite{CDF}. The measured branching ratio is comparable with the SM prediction \cite{Aliev} within the errors 
of form factors. Comparing the different NP models' predictions with the experimental data on this channel, it is possible to obtain information about and put 
limits on the parameters of the models. In our previous work, we put a lower limit to the compactification parameter of the universal extra dimension (UED) via 
this channel comparing the theoretical calculations with the experimental data \cite{azizi}.  

The LHCb experiment at the LHC  has been 
taking data for proton-proton collision in 2011 and 2012 at $\sqrt{s}=7$ and $8$ TeV,  respectively, integrating a luminosity in excess of $3 fb^{-1}$ \cite{LHCb1,LHCb2}. 
The LHCb measurement on the differential branching ratio of the
  $\Lambda_b \rightarrow \Lambda \mu^+\mu^-$ is in its final stage \cite{LHCb}. Considering these experimental progresses and the accessed luminosity we hope
we will able to study more decay channels such as the radiative baryonic decay of $\Lambda_b \rightarrow \Lambda \gamma$ at
 LHCb \cite{LHCb,LHCb1,LHCb2,Mancinelli}. 
  In this connection, we study this radiative  
decay channel in SM as well as UED with a single ED (UED5) and  two EDs (UED6) in the present work. There are many works dedicated to the analysis of different 
decay channels in UED5 in the literature (for some of them see \cite{azizi,azizi2,Buras1,KK,Colangelo2,Bashiry,nihan,Bayar,Aslam,Aliev-Savci,De-Fazio,Sirvanli-azizi,Pak,Aliev-Sirvanli,Aslam2,Colangelo,Giri}).
 However, the number of works devoted to the applications of the  UED6 is relatively few. As the expression of the only Wilson coefficient $C^{eff}_{7}$ now is available in UED6 \cite{Freitas2}, 
it is possible to study the radiative channels based on the $b \rightarrow s \gamma$.

 In \cite{Biancofiore-Fazio}, the  UED6 is employed to analyze the $B\to K \eta^{(\prime)} \gamma$ decay channel, where by comparing the results with the 
experimental data, a lower limit of  $400$ GeV is put for the compactification scale.
 For some other previous constraints on the compactification factor obtained via electroweak 
precision tests, some cosmological constraints and different hadronic channels in UED5 see for instance \cite{azizi,Gogoladze,Cembranos,Colangelo2,Agashe,ACD,Appelquist-Yee}. We shall use the latest  
lower limits on the compactification factor $1/R$ obtained from different FCNC transitions in UED5 model \cite{Haisch},  some FCNC transitions in UED6 model \cite{Biancofiore-Fazio}, 
electroweak precision tests \cite{Gogoladze}, 
  cosmological constraints \cite{Cosm-Belanger}, direct searches \cite{ATLAS} as well as  the latest results of the Higgs search at the LHC and of the electroweak precision data for the
S and T parameters \cite{takuya}.

 Scenarios with EDs play crucial roles among models beyond the SM. The main feature that leads to the difference among ED models is the number of dimensions added to the SM. 
In the UED5, we have an extra universal compactified dimension compared to the SM, while in UED6 we consider two extra UEDs.  Because of the universality, the SM particles 
can propagate into the UEDs and interact with the Kaluza-Klein (KK) modes existing in EDs. As a result of these interactions, the new Feynman diagrams appear and this leads 
to a modification in the Wilson coefficients entered the low energy Hamiltonians defining the hadronic decay channels \cite{Buras,Buras2,KK,Freitas2}. 
In the UED5, the ED is compactified to the orbifold $S^1/Z_2$, with the fifth coordinate $x_5=y$ changing from $0$ to $2 \pi R$. The points $y=0$ and $y=\pi R$ are fixed points
 of this orbifold. The boundary conditions at these points give the KK mode expansion of the fields. The masses of the KK particles in this model are obtained in terms of compactification
 scale as ${m_n}^2={m_0}^2+n^2/R^2$  where $n=1,2,...$ and $m_0$ represents the zeroth mode mass referring to the SM particles (for more about the model 
see \cite{KK,Appelquist-Yee,ACD,Antoniadis,Antoniadis2,Arkani,Arkani1,Randall1,Randall2}). 
  
Models with two EDs  are more attractive since they reply to some questions existing in the SM \cite{Freitas}. In this model, cancellations of chiral anomalies
 allow the existence of the right-handed neutrinos and predict the correct number of the fermion families \cite{Freitas,Poppitz,Burdman}. At the same time, this model 
provides a natural explanation for the long lifetime of the proton \cite{Ghosh,Ponton}. In  UED6 models also, all the  SM fields  are assumed to propagate into both flat
 EDs that are already compactified on  a chiral square of the side $L=\pi R$ \cite{Freitas,Dobrescu,Freitas2}. The KK particles existing in this  model are marked by two
 positive integers $k$ and $l$ which symbolize quantization of momentum along the EDs. The  masses of these particles are given in terms of the compactification scale 
by $M_{(k,l)}=\sqrt{k^{2}+l^{2}}/R$ \cite{Freitas}. In this model, particles on first KK level with KK numbers ($1,0$)  are odd under KK parity. These particles may be 
produced only in pairs at colliders. The  particles on level-$2$ are even under KK parity and have KK numbers ($1,1$) \cite{Dobrescu}. This may lead to a totally 
different sets of signatures involving the resonances of the heavy top and bottom quarks \cite{Dobrescu,Dobrescu2}. The masses of particles on level-2 are $\sqrt{2}$ 
factor larger than the masses of particles on level-1 \cite{Freitas}. This makes the particles at level -2 be most easily accessible at LHC \cite{Dobrescu2}. For more 
details about the UED6 model and some of its applications see for instance \cite{Freitas2,Freitas,Burdman,Ghosh,Ponton,Dobrescu2}. 

The outline of the article is  as follows. In next section, we present the effective  Hamiltonian responsible for the  $\Lambda_b \rightarrow \Lambda \gamma$ 
in SM, UED5 and UED6 as well as the transition matrix elements in terms of form factors. In section 3, we calculate the decay width and branching ratio of the 
decay under consideration and numerically analyze them. In this section, we also compare the results of UED5 and UED6 with the SM predictions and look for the 
deviations from the SM at different values of the compactification radius.
\section{The radiative $\Lambda_b \rightarrow \Lambda \gamma$ transition in SM, UED5 and UED6 models}       
In the present section, we present the effective Hamiltonian and show how the Wilson coefficient $C^{eff}_{7}$ changes both in UED scenarios with one and two extra dimensions compared to the SM. We also define the transition matrix elements appeared in the amplitude of the considered decay in terms of form factors.
\subsection{The effective Hamiltonian}
 At quark level, the general effective Hamiltonian for $b \rightarrow s \gamma$ and $b \rightarrow s g$ transitions  in SM  and in terms of Wilson coefficients and operators is given by \cite{Buras}
\bea \label{Heff}
 {\cal H}^{eff}&=& -\frac{G_{F}}{\sqrt{2}}V_{tb}V_{ts}^\ast 
\bigg[ {\sum\limits_{i=1}^{6}} C_{i}({\mu}) Q_{i}({\mu})+C_{7
\gamma}(\mu) Q_{7\gamma}(\mu)+C_{8G}(\mu) Q_{8G}(\mu)\bigg],
\eea
where $G_F$ is the Fermi weak coupling constant and $V_{ij}$ are elements of the Cabibbo-Kobayashi-Maskawa (CKM) mixing matrix. The complete list of the operators entered to the above Hamiltonian is given as
\bea
  Q_{1}~\es ({\bar{s}}_{\alpha}c_{\beta} )_{V-A}
               ({\bar{c}}_{\beta} b_{\alpha})_{V-A},~\nnb \\
  Q_{2}~\es({\bar{s}}_{\alpha}c_{\alpha})_{V-A}
               ({\bar{c}}_{\beta} b_{\beta} )_{V-A},~\nnb \\
  Q_{3}~\es({\bar{s}}_{\alpha}b_{\alpha})_{V-A}\sum\limits_{q}
           ({\bar{q}}_{\beta} q_{\beta} )_{V-A},~\nnb \\
  Q_{4}~\es({\bar{s}}_{\beta}b_{\alpha})_{V-A}\sum\limits_{q}
           ({\bar{q}}_{\alpha} q_{\beta} )_{V-A},~\nnb \\
  Q_{5}~\es({\bar{s}}_{\alpha}b_{\alpha})_{V-A}\sum\limits_{q}
           ({\bar{q}}_{\beta} q_{\beta} )_{V+A},~\nnb \\
  Q_{6}~\es({\bar{s}}_{\beta}b_{\alpha})_{V-A}\sum\limits_{q}
           ({\bar{q}}_{\alpha} q_{\beta} )_{V+A},~\nnb \\
  Q_{7 \gamma}\es{e \over 4\pi^2} \bar{s}_{\alpha}\sigma^{\mu \nu}(m_b R+m_sL)b_{\alpha}\, F_{\mu \nu},~\nnb \\
  Q_{8 G}\es {g_s \over 4\pi^2} \bar{s}_{\alpha}\sigma^{\mu \nu}(m_b R+m_sL)T^a_{\alpha \beta}b_{\beta}\,
    G^{a}_{\mu \nu },~ \eea
where $Q_{1,2}$, $Q_{3,4,5,6}$ and $Q_{7 \gamma,8 G}$ are the current-current (tree), QCD penguin and the magnetic penguin operators, respectively.  $\alpha$  and $\beta$  are the color indices, $R=(1+\gamma_5)/2$ is the right-handed projector and $L=(1-\gamma_5)/2$ is the left-handed projector. In the above operators $e$ and $g_s$ are the coupling constants of the electromagnetic and strong interactions, respectively. $F_{\mu\nu}$ is the field strength tensor of the electromagnetic field and is defined by
\bea \label{emtensor} F_{\mu\nu}(x) &=& -i(\varepsilon_{\mu}q_{\nu}-\varepsilon_{\nu}q_{\mu})e^{iqx}~, \eea
where $\varepsilon_{\mu}$ is the polarization vector of the photon and $q$ is its momentum. 
The most relevant contribution to $b \rightarrow s \gamma$ comes from the magnetic penguin operator $Q_{7 \gamma}$. Hence the effective Hamiltonian in our case can be written as,
\bea \label{Heff} {\cal H}^{eff}(b \rightarrow s \gamma) &=& -{G_F e \over 4\pi^2\sqrt{2}} V_{tb}
V_{ts}^\ast  C^{eff}_{7}(\mu)\bar{s} \sigma_{\mu\nu} \Big[m_{b}R+m_{s}L \Big]bF^{\mu\nu}~, \eea
where $C^{eff}_{7}$ is relevant the Wilson coefficient. 
 Under scenarios with EDs including one or two compact extra dimensions, the form of effective Hamiltonian remains unchanged, but the Wilson coefficient $C^{eff}_{7}$  is modified because of  additional Feynman diagrams  coming from the interactions of the  KK particles with themselves as well as the SM particles in the bulk.
 This coefficient in SM is given as \cite{C7eff}
 \bea
\label{wilson-C7eff} C_7^{eff}(\mu_b) \es
\eta^{\frac{16}{23}} C_7(\mu_W)+ \frac{8}{3} \left(
\eta^{\frac{14}{23}} -\eta^{\frac{16}{23}} \right) C_8(\mu_W)+C_2 (\mu_W) \sum_{i=1}^8 h_i \eta^{a_i}~, \nnb\\ \eea
 where
 \bea \eta \es
\frac{\alpha_s(\mu_W)} {\alpha_s(\mu_b)}~,\eea
and
\bea
\alpha_s(x)=\frac{\alpha_s(m_Z)}{1-\beta_0\frac{\alpha_s(m_Z)}{2\pi}\ln(\frac{m_Z}{x})}.\eea
Here $\alpha_s(m_Z)=0.118$ and $\beta_0=\frac{23}{3}$.
The values of coefficients $a_i$ and $h_i$ in Eq.\eqref{wilson-C7eff} are
given as
\be\frac{}{}
   \label{coefficients}
\begin{array}{rrrrrrrrrl}
a_i = (\!\! & \f{14}{23}, & \f{16}{23}, & \f{6}{23}, & -
\f{12}{23}, &
0.4086, & -0.4230, & -0.8994, & 0.1456 & \!\!)  \vspace{0.1cm},\\
h_i = (\!\! & 2.2996, & - 1.0880, & - \f{3}{7}, & - \f{1}{14}, &
-0.6494, & -0.0380, & -0.0186, & -0.0057 & \!\!). 
\end{array}
\ee
Also  $C_2(\mu_W)$, $C_7(\mu_W)$ and $C_8(\mu_W)$ in Eq.\eqref{wilson-C7eff} are defined in the following way:
\bea
 C_2(\mu_W)=1~,~~ C_7(\mu_W)=-\frac{1}{2}
D^\prime_0(x_t)~,~~ C_8(\mu_W)=-\frac{1}{2}
E^\prime_0(x_t)~ , \eea 
where  $D^\prime_0(x_t)$ and $E^\prime_0(x_t)$ 
are expressed as \bea \label{Dprime0} D^\prime_0(x_t) \es -
\frac{(8 x_t^3+5 x_t^2-7 x_t)}{12 (1-x_t)^3}
+ \frac{x_t^2(2-3 x_t)}{2(1-x_t)^4}\ln x_t~, \\ \nnb \\
\label{Eprime0} E^\prime_0(x_t) \es - \frac{x_t(x_t^2-5 x_t-2)}
{4(1-x_t)^3} + \frac{3 x_t^2}{2 (1-x_t)^4}\ln x_t~. \eea 
The Wilson coefficient $C^{eff}_{7}$  in  UED5 has been calculated in \cite{Buras,Buras1,KK,Misiak,Munz,C7eff}. In this model, each periodic function $F(x_t,1/R)$ ($F=$ $D^\prime$ or $E^\prime$) inside the Wilson coefficient includes   a SM part $F_0(x_t)$ plus an additional part in terms of compactification factor $1/R$ due to new interactions, i.e.,
\bea F(x_t,1/R)=F_0(x_t)+\sum_{n=1}^{\infty}F_n(x_t,x_n),\label{function} \eea
where $x_{t}=\frac{m_{t}^{2}}{m_{W}^{2}}$,   $x_n=\displaystyle{m_n^2 \over m_W^2}$, and  $m_n=\displaystyle{n \over R} $.  Here $m_t$,  $m_{W}$ and $m_n$ are  masses of the top quark,  $W$ boson and KK particles (non-zero modes), respectively. In UED5, the functions $D^\prime(x_t,1/R)$ and $E^\prime(x_t,1/R)$ in terms of compactification parameter $1/R$ are given as 
 \bea D^\prime
(x_t,1/R)=D^\prime_0(x_t)+\sum_{n=1}^{\infty}D^\prime_n(x_t,x_n),~~~~~E^\prime
(x_t,1/R)=E^\prime_0(x_t)+\sum_{n=1}^{\infty}E^\prime_n(x_t,x_n)~,
\eea 
where the functions including KK contributions are written as
 \bea
\label{DN-prime} \sum_{n=1}^{\infty}D^\prime_n(x_t,x_n) \es
\frac{x_t[37 - x_t(44+17 x_t)]}{72 (x_t-1)^3} \nnb \\
\ar \frac{\pi m_W R}{12} \Bigg[ \int_0^1 dy \, (2 y^{1/2}+7
y^{3/2}+3 y^{5/2}) \, \coth (\pi m_WR \sqrt{y}) \nnb \\
\ek \frac{x_t (2-3 x_t) (1+3 x_t)}{(x_t-1)^4}J(R,-1/2)\nnb \\
\ek \frac{1}{(x_t-1)^4} \{ x_t(1+3 x_t)+(2-3 x_t)
[1-(10-x_t)x_t] \} J(R, 1/2)\nnb \\
\ek \frac{1}{(x_t-1)^4} [ (2-3 x_t)(3+x_t) + 1 - (10-x_t) x_t]
J(R, 3/2)\nnb \\
\ek \frac{(3+x_t)}{(x_t-1)^4} J(R,5/2) \Bigg]~, \nnb\\
\eea 
and
\bea \label{EN-prime}
\sum_{n=1}^{\infty}E^\prime_n(x_t,x_n)\es
\frac{x_t[17+(8-x_t)x_t]}
{24 (x_t-1)^3} \nnb \\
\ar \frac{\pi m_W R}{4} \Bigg[\int_0^1 dy \, (y^{1/2}+
2 y^{3/2}-3 y^{5/2}) \, \coth (\pi m_WR \sqrt{y}) \nnb \\
\ek {x_t(1+3 x_t) \over (x_t-1)^4}J(R,-1/2)\nnb \\
\ar \frac{1}{(x_t-1)^4} [ x_t(1+3 x_t) - 1 + (10-x_t)x_t] J(R, 1/2)\nnb \\
\ek \frac{1}{(x_t-1)^4} [(3+x_t)-1+(10-x_t)x_t ]J(R, 3/2)\nnb \\
\ar{(3+x_t) \over  (x_t-1)^4} J(R,5/2)\Bigg]~, \eea
where 
\bea
\label{J} J(R,\alpha)=\int_0^1 dy \, y^\alpha \left[ \coth
(\pi m_W R \sqrt{y})-x_t^{1+\alpha} \coth(\pi m_t R \sqrt{y})
\right]~. \eea
 %
 The  Wilson coefficient $C_7^{eff}(1/R)$  in the UED6 model with two extra dimensions is given by \cite{Freitas2}
\bea
C^{\rm eff}_i (\mu) = C^{\rm eff}_{i \, {\rm SM}} (\mu) + \Delta
C^{\rm eff}_i (\mu) \, , \quad i = 1, \ldots, 8 
\eea
where
\bea
\Delta C^{\rm eff}_i (\mu) = \sum_{n = 0}^{\infty} \left (
  \f{\alpha_{s}}{4 \pi} \right )^n \Delta C_i^{{\rm eff} (n)} (\mu) \, , 
\eea
and
\bea 
\Delta C_i^{{\rm eff} (0)} (\mu_{0}) = 
\begin{cases}
  \phantom{-}0 & \mbox{for $i = 1, \ldots , 6$,} \\[1ex] 
  -\f{1}{2} {\sum\limits_{k, l}}^{\prime} A^{(0)} (x_{kl}) &
  \mbox{for $i = 7$,} \\[2ex] 
  -\f{1}{2} {\sum\limits_{k, l}}^{\prime} F^{(0)} (x_{kl}) & 
  \mbox{for $i = 8$.} 
\end{cases}
\eea
The superscript ($^\prime$) in summation means that the KK sums run only over the restricted ranges $k \geq 1$ and $l \geq 0$,  i.e.,  $\sum_{k, l}^{\prime} = \sum_{k\geq 1} \sum_{l\geq 0}$. 
The upper limits for $k$ and $l$ are restricted as $k+l \leq N_{KK}$ where $N_{KK}$ can get values in the interval $(5-15)$ \cite{Freitas2}.
The parameter $N_{KK}$ in our calculations is the total number of contributing KK modes \cite{ACD}. 
The highest KK level in this compactification is fixed by $N_{KK}=\Lambda R$ \cite{Chivukula}, where $\Lambda$ is a  scale at which the QCD interactions become strong in the ultraviolet \cite{Dobrescu2}.   
In the case of UED5 $N_{KK}=n$, however, as the KK sums over $n$ up to infinity  is convergent we have no dependence on 
the $N_{KK}$ after the  KK sums. In the case of UED6 the KK mode sums diverge in the limit $N_{KK}\rightarrow \infty$ because the KK spectrum is denser than the UED5 case. 
The electroweak observables convergence in four and five dimensions at one loop, become logarithmically divergent at $d=6$ and more divergent in higher dimensions \cite{ACD}. Hence we should put
 a cut-off and, as a result, an upper limit to $k+l$.

 The Inami-Lim functions inside the $C_7^{eff}$ in leading order are decomposed as  
\bea 
X^{(0)} (x_{kl}) & =\sum\limits_{I = {\scriptstyle W}, a, H} X^{(0)}_I (x_{kl}) \, , \quad X = A, F \, 
\eea
where $x_{kl}$ is defined as
\bea
x_{kl} = (k^2+l^2)/(R^2 m_{W}^2),
\eea
and the functions $ X^{(0)}_{W, a, H} (x_{kl}) $
define the contributions because of the exchange of KK modes which
would be the Goldstone bosons $G^\pm_{(kl)}$, $W$-bosons $W^\pm_{\mu(kl)}$ and the scalar fields $a^\pm_{(kl)}$ as well as $W^\pm_{H (kl)}$. They are given as 
\bea 
A_{W}^{(0)} (x_{kl}) &=&  \frac{x_t(6((x_t-3) x_t+3) x_{kl}^2-3 (5 (x_t-3) x_t+6) x_{kl}+x_t (8
x_t+5)-7)}{12 (x_t-1)^3} \nonumber \\
 &+& \f{1}{2} (x_{kl}-2)x_{kl}^2 \ln \left(\f{x_{kl}}{x_{kl}+1}\right)-\frac{(x_{kl}+x_t)^2 (x_{kl}+3 x_t-2)}{2 (x_t-1)^4}\ln \left(\frac{x_{kl}+x_t}{x_{kl}+1}\right)~, \nnb\\ \eea
\bea
F_{W}^{(0)} (x_{kl}) &=& \frac{x_t \left(-6((x_t-3) x_t+3) x_{kl}^2-3 ((x_t-3) x_t+6) x_{kl}+(x_t-5) x_t-2\right)}{4 (x_t-1)^3} \nonumber \\
 &-&\f{3}{2} (x_{kl}+1)x_{kl}^2 \ln \left(\frac{x_{kl}}{x_{kl}+1}\right) +\frac{3 (x_{kl}+1) (x_{kl}+x_t)^2}{2 (x_t-1)^4} \ln \left(\frac{x_{kl}+x_t}{x_{kl}+1}\right)~, \nnb\\ \eea
\bea
A_a^{(0)} (x_{kl}) &=& \frac{x_t \left(6 x_{kl}^2-3 (x_t (2 x_t-9)+3) x_{kl}+(29-7 x_t) x_t-16\right)}{36 (x_t-1)^3}\nonumber \\
&-& \frac{(x_{kl}+3 x_t-2) (x_t+x_{kl} ((x_{kl}-x_t+4) x_t-1)) }{6 (x_t-1)^4} \ln \left(\f{x_{kl}+x_t}{x_{kl}+1}\right)\nonumber \\
&-& \frac{1}{6} (x_{kl}-2) x_{kl} \ln \left(\frac{x_{kl}}{x_{kl}+1}\right) ~, \nnb\\ \eea
\bea
F_a^{(0)} (x_{kl}) &=& \frac{x_t \left(-6 x_{kl}^2+\left(6 x_t^2-9 x_t-9\right) x_{kl}+(7-2 x_t) x_t-11\right)}{12 (x_t-1)^3}\nonumber \\
&+& \frac{(x_{kl}+1) (x_t+x_{kl} ((x_{kl}-x_t+4) x_t-1))}{2 (x_t-1)^4} \ln \left(\frac{x_{kl}+x_t}{x_{kl}+1}\right)\nonumber \\
&+& \frac{1}{2} x_{kl} (x_{kl}+1) \ln \left(\f{x_{kl}}{x_{kl}+1}\right) ~, \nnb\\ \eea
\bea
A_H^{(0)} (x_{kl}) & = & \frac{x_t \left(6 \left(x_t^2-3 x_t+3\right) x_{kl}^2-3 \left(3 x_t^2-9 x_t+2\right) x_{kl}-7 x_t^2+29 x_t-16\right)}{36 (x_t-1)^3} \nonumber \\
&-&\frac{(x_{kl}+1) \left(x_{kl}^2+(4 x_t-2) x_{kl}+x_t (3 x_t-2)\right)}{6(x_t-1)^4} \ln \left(\frac{x_{kl}+x_t}{x_{kl}+1}\right)\nonumber \\
&+&\frac{1}{6} x_{kl} \left(x_{kl}^2-x_{kl}-2\right) \ln \left(\frac{x_{kl}}{x_{kl}+1}\right)~, \nnb\\ \eea
and
\bea
F_H^{(0)} (x_{kl})& = & -\frac{x_t \left(6 \left(x_t^2-3 x_t+3\right) x_{kl}^2+3 \left(3 x_t^2-9 x_t+10\right) x_{kl}+2 x_t^2-7 x_t+11\right)}{12 (x_t-1)^3} \nonumber \\ 
& - & \frac{1}{2} x_{kl} (x_{kl}+1)^2 \ln \left(\frac{x_{kl}}{x_{kl}+1}\right) +\frac{(x_{kl}+x_t)
(x_{kl}+1)^2}{2 (x_t-1)^4} \ln \left(\frac{x_{kl}+x_t}{x_{kl}+1}\right)~. \nnb\\ \eea 
 %
 %
 %
 %
 %
\subsection{Transition amplitude and  matrix elements}
 The amplitude for this transition is obtained by sandwiching the effective Hamiltonian between the final and initial baryonic states 
\bea \label{Amplitude} {\cal M}^{(\Lambda_b
\rightarrow \Lambda \gamma)}&=& \langle \Lambda(p_{\Lambda})|{\cal H}^{eff}|\Lambda_{b}(p_{\Lambda_{b}})\rangle ~, \eea
where $p_{\Lambda}$ and $p_{\Lambda_{b}}$ are momenta of the $\Lambda$ and $\Lambda_{b}$ baryons, respectively. In order to proceed, 
we need to define the following transition matrix elements in terms of two form factors $f_2^{T}$ and $g_2^{T}$:
\begin{eqnarray} \label{Transition-matrix-element}
\langle \Lambda(p_{\Lambda})|\bar s\;\sigma_{\mu\nu}q^\nu ( g_V &+& \gamma_5g_A ) b|\Lambda_{b}(p_{\Lambda_{b}})\rangle \nnb\\&=& \bar
u_{\Lambda}(p_{\Lambda})\sigma_{\mu\nu}q^\nu \Big(g_V f_2^{T}(0)+\gamma_5 g_A
g_2^{T}(0) \Big)u_{\Lambda_{b}}(p_{\Lambda_{b}}),
\end{eqnarray}
where $g_V=1+m_{s}/m_{b}$, $g_A=1-m_{s}/m_{b}$, and $\bar u_{\Lambda}$ and $u_{\Lambda_{b}}$ are spinors of the $\Lambda$ and $\Lambda_{b}$ baryons, respectively. In the following, we will use the values of the form factors calculated via light cone QCD sum rules in full theory \cite{Aliev}. 
\section{Decay width and branching ratio}
In this section we would like to calculate the total decay width and branching ratio of the transition under consideration. Using the aforesaid transition matrix elements in terms of form factors, 
we find the $1/R$-dependent total decay width in terms of the two form factors as
\bea \label{DiffDecayRate}
\Gamma_{(\Lambda_b\rightarrow\Lambda\gamma)}(1/R)=\frac{G_F^2 \alpha_{em}|V_{tb}V_{ts}^*|^2 m_b^2}{64\pi^4}|C_7^{eff}(1/R)|^2 \left(\frac{m_{\Lambda_b}^2-m_\Lambda^2}{m_{\Lambda_b}} \right)^3\Big(g_V^2
|f_2^{T}(0)|^2 + g_A^2 |g_2^{T}(0)|^2\Big),\nnb\\ \eea
where $\alpha_{em}$ is the fine structure constant at Z mass scale. In order to calculate the $1/R$-dependent branching ratio, we need to multiply the total decay width  by the
 lifetime of the initial baryon $\Lambda_b$ and divide by $\hbar$. To numerically analyze the obtained results,  we use some input parameters as presented in Table 1. 
For the quark masses, we use the $\overline{MS}$ scheme  values \cite{PDG} (see Table 2).
\begin{table}[ht]
\centering
\rowcolors{1}{lightgray}{white}
\begin{tabular}{cc}
\hline \hline
   Input Parameters  &  Values    
           \\
\hline \hline
$ m_{W} $            &   $ 80.38     $ $GeV$             \\
$ m_{\Lambda_b} $    &   $ 5.619    $ $GeV$               \\
$ m_{\Lambda} $      &   $ 1.1156   $ $GeV$                \\
$ \mu_{b} $          &   $ 5        $ $GeV$                 \\
$ \mu_{W} $          &   $ 80.4     $ $GeV$                  \\
$ \mu_{0} $          &   $ 160      $ $GeV$                   \\
$ \tau_{\Lambda_b} $ &   $ 1.425\times 10^{-12} $ $s$          \\
$ \hbar  $           &   $ 6.582\times 10^{-25} GeV s $         \\
$ G_{F} $            &   $ 1.17\times 10^{-5} $ $GeV^{-2}$       \\
$ \alpha_{em} $      &   $ 1/137 $                                \\
$ | V_{tb}V_{ts}^*|$ &   $ 0.041 $                                 \\
 \hline \hline
\end{tabular}
\caption{The values of some input parameters, mainly taken from PDG \cite{PDG}, ​​used in the numerical analysis.}
\end{table}
\begin{table}[ht]
\centering
\rowcolors{1}{lightgray}{white}
\begin{tabular}{cc}
\hline \hline
   Quarks   &  masses in  $\overline{MS}$ scheme    
           \\
\hline \hline
$ m_{s} $            &     $ (0.095 \pm0.005)   $ $GeV$      \\
$ m_{b} $            &    $ (4.18 \pm0.03)    $ $GeV$     \\
$ m_{t} $            &    $ 160^{+4.8}_{-4.3}      $ $GeV$      \\
\hline \hline
\end{tabular}
\caption{The values of  quark masses in  $\overline{MS}$ scheme \cite{PDG}.}
\end{table}

As we previously mentioned, we use the values of form factors calculated via light cone QCD sum rules in full theory as the main inputs in numerical analysis \cite{Aliev}. Their values are presented in Table 3.
\begin{table}[ht]
\centering 
\rowcolors{1}{lightgray}{white}
\begin{tabular}{ccc}
\hline \hline
            & \mbox{form factors at} $q^2=0$ \\
          \hline \hline
$ f_2^{T}(0) $  &  $0.295 \pm 0.105 $      \\
$ g_2^{T}(0) $  &  $0.294 \pm 0.105$       \\
 \hline \hline
 \end{tabular}
\caption{The values of form factors $f^T_{2}(0)$ and  $g^T_{2}(0)$ \cite{Aliev}.}
\end{table}

In this part we present the numerical values of the Wilson coefficient $C^{eff}_{7}$ obtained from the previously presented formulas in SM, UED5 and UED6 models.
 In SM, its value is obtained as  $C^{eff}_{7}=-0.295$. 
We depict the values of the Wilson coefficient $C^{eff}_{7}$ at different values of $1/R$  in UED5 and UED6 scenarios with  $N_{KK}=(5,10,15)$ in Table 4.
\begin{table}[ht]
\centering
\rowcolors{1}{lightgray}{white}
\begin{tabular}{ccccc}
\hline \hline
   {\footnotesize $1/R$}  & {\footnotesize $ C^{eff}_{7}$}  & {\footnotesize $ C^{eff}_{7}$}  &  {\footnotesize $C^{eff}_{7}$ } &  {\footnotesize $C^{eff}_{7}$ }
                          \\ 
     {\footnotesize [GeV] }   &  {\footnotesize (UED5)}  &   {\footnotesize (UED6 for $N_{KK}=5$)} & {\footnotesize (UED6 for $N_{KK}=10$)}  & {\footnotesize (UED6 for $N_{KK}=15$)}     \\
\hline \hline
$ 200 $    &  $ -0.198  $    &   $ -0.053  $   &  $  0.048  $  & $ 0.110   $ \\
$ 400 $    &  $ -0.265  $    &   $ -0.224  $   &  $ -0.198  $  & $ -0.182  $  \\
$ 600 $    &  $ -0.281  $    &   $ -0.262  $   &  $ -0.250  $  & $ -0.243  $   \\
$ 800 $    &  $ -0.287  $    &   $ -0.276  $   &  $ -0.269  $  & $ -0.265  $    \\
$ 1000$    &  $ -0.289  $    &   $ -0.283  $   &  $ -0.278  $  & $ -0.279  $     \\
 \hline \hline
\end{tabular}
\caption{The numerical values of Wilson coefficient $C^{eff}_{7}$ at the different values of $1/R$ in UED5 and UED6  for $N_{KK}=(5,10,15)$.}
\end{table}

Making use of all given input values we find the value of the branching ratio  in SM  as presented in Table 5. For comparison, we also give the results of other 
related works \cite{HeLiWang,Colangelo3,Huang,Mohanta,GanLiu,Mannel} in 
the same Table  as well as the upper limit from PDG \cite{PDG}. 
From this Table we see that, within the errors, our result is consistent with those  of QCD sum rules \cite{Colangelo3,Huang} and CZ current \cite{GanLiu} and exactly 
the same  with pole model's prediction \cite{Mannel}. However, our prediction  differs considerably from these of light cone QCD sum rules \cite{HeLiWang}, covariant 
oscillator quark model (COQM) \cite{Mohanta} and Ioffe current \cite{GanLiu}. The difference between our SM prediction on the branching ratio with that of \cite{HeLiWang} with the same method can be attributed
to the point that in \cite{HeLiWang} the authors  consider the distribution amplitudes (DAs) of $\Lambda$ baryon as the main inputs of the light cone QCD sum rule method up to twist 6, however, in our case the form factors
 have been calculated considering the DAs up 
to twist 8. Besides, in \cite{HeLiWang} the higher conformal spin contributions to the DAs are not taken into account, while the calculations of form factors in our case include these contributions. Finally,  
in \cite{HeLiWang} the form factors are calculated in heavy quark effective limit while we use form factors calculated in full QCD without any approximation. 
The order of branching ratio shows that this channel can be accessible at LHCb. 
\begin{table}[ht]
\centering
\rowcolors{1}{lightgray}{white}
\begin{tabular}{ccc}
\hline \hline
     \mbox{Ref.}  & $\textit{BR}$($\Lambda_b \rightarrow \Lambda \gamma$) \\
          \hline \hline
 Our result             &  $(1.003-4.457) \times 10^{-5}         $       \\
 Light-cone sum rule \cite{HeLiWang} &  $(0.63-0.73) \times 10^{-5}$              \\
 Three-point QCD sum rule \cite{Colangelo3}  & $(3.1 \pm 0.6)  \times 10^{-5}$    \\
 QCD sum rule \cite{Huang}            &  $(3.7 \pm 0.5)  \times 10^{-5}$             \\
 COQM \cite{Mohanta}                 &  $0.23 \times 10^{-5}           $           \\
 CZ current \cite{GanLiu}       &  ($1.99^{+0.34}_{-0.31}$)$ \times 10^{-5}$     \\
 Ioffe current \cite{GanLiu}    &  ($0.61^{+0.14}_{-0.13}$)$ \times 10^{-6}$      \\
 Pole Model \cite{Mannel}        & $(1.0-4.5) \times 10^{-5}$     \\
 PDG \cite{PDG}                    & $ < 1.3 \times 10^{-3} $ ($CL=90\%)  $ \\
          \hline \hline
      \end{tabular}
       \caption{The values of branching ratio in SM.}
        \end{table}

In order to look for  the differences between the predictions of the SM and the considered UED scenarios, we present the dependence of the central values of the branching ratio on $1/R$ at different 
models in figure 1. Note that to better see the deviations between the SM predictions and those of UED scenarios, in all figures,  we plot the branching ratio in terms of $1/R$ in the interval
 $200~GeV\leq1/R\leq2000~GeV$. However we will consider the latest lower limits on the compactification factor obtained from different approaches in our analysis and discussions.
The latest  lower limits on $1/R$  are: $400~GeV$  put by some FCNC transitions in UED6 model \cite{Biancofiore-Fazio},
 $500~GeV$  put via cosmological constraints \cite{Cosm-Belanger}, $600~GeV$  obtained via different FCNC transitions in UED5 model (for instance see \cite{Haisch}) and 
electroweak precision tests \cite{Gogoladze}, $1.41~TeV$  quoted via direct searches at ATLAS Collaboration \cite{ATLAS} as well as $650~(850\sim1350)~GeV$ from the latest results of the Higgs search/discovery at the LHC
for UED5 (UED6) \cite{takuya} and $700~(900\sim1500)~GeV$ from the electroweak precision data for S and T parameters in the case of UED5 (UED6) \cite{takuya}.

\begin{figure}[h!]
\centering
\begin{tabular}{cc}
\epsfig{file=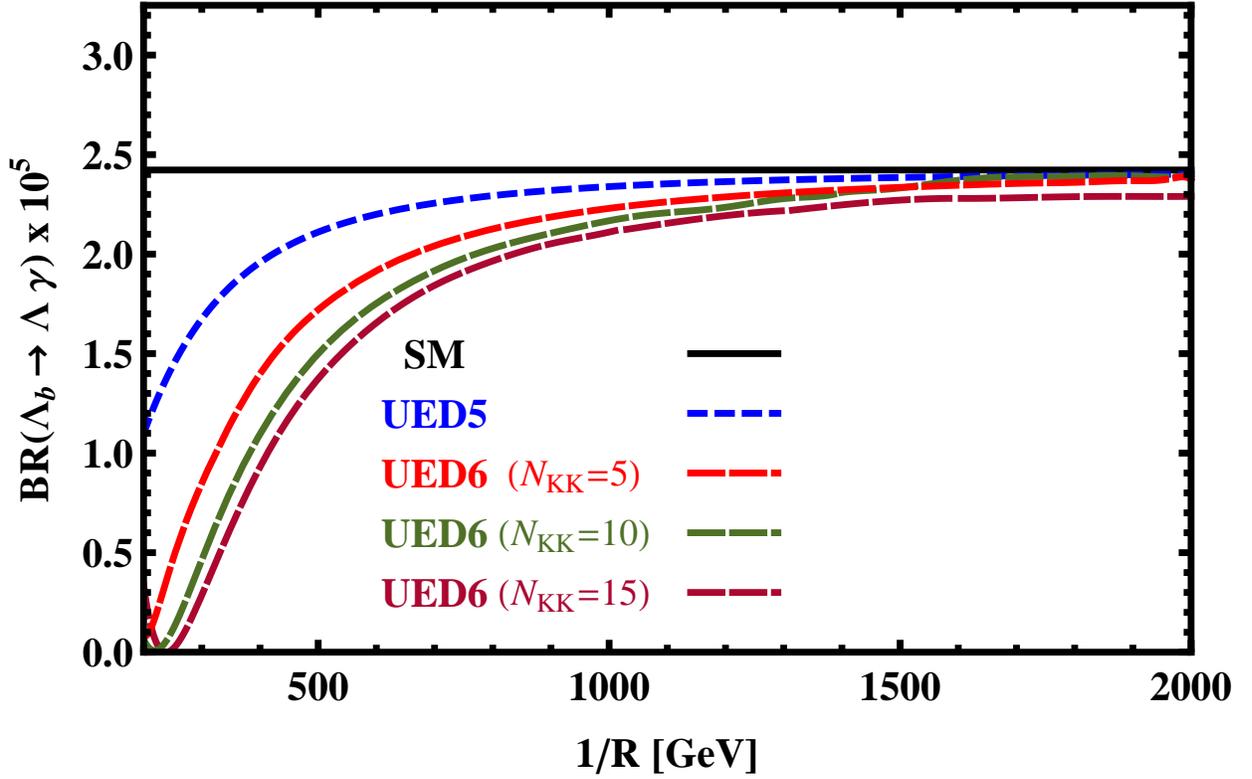,width=1.0\linewidth,clip=}
\end{tabular}
\caption{The dependence of branching ratio for $\Lambda_b
\rightarrow \Lambda \gamma$ decay channel on compactification factor $1/R$ in SM, UED5 and UED6 models with $N_{KK}=(5,10,15)$ when the central values of the form factors are used. }
\end{figure}
From  figure 1 we see that
there are distinctive differences between the SM predictions and those of UED models, especially UED6 for $N_{KK}=15$, at small values of the compactification factor $1/R$. These differences exist in the
 lower limits obtained by   different FCNC transitions in UED5 and UED6, cosmological constraints, electroweak precision tests \cite{Biancofiore-Fazio,Cosm-Belanger,Haisch,Gogoladze} and the latest results of the Higgs search at the LHC and of the electroweak precision data for the
S and T parameters \cite{takuya}, however, they become small
when $1/R$ approaches to $1~TeV$. Our analysis show that the UED scenarios give close  results to the SM for $1/R\geq1~TeV$. Hence, when considering the lower limit $1.41~TeV$ 
 quoted via direct searches at ATLAS Collaboration \cite{ATLAS} we see very small deviations of the UED models predictions from those of the SM for the decay channel under consideration.
%
%

\begin{figure}[h!]
\centering
\begin{tabular}{ccc}
\epsfig{file=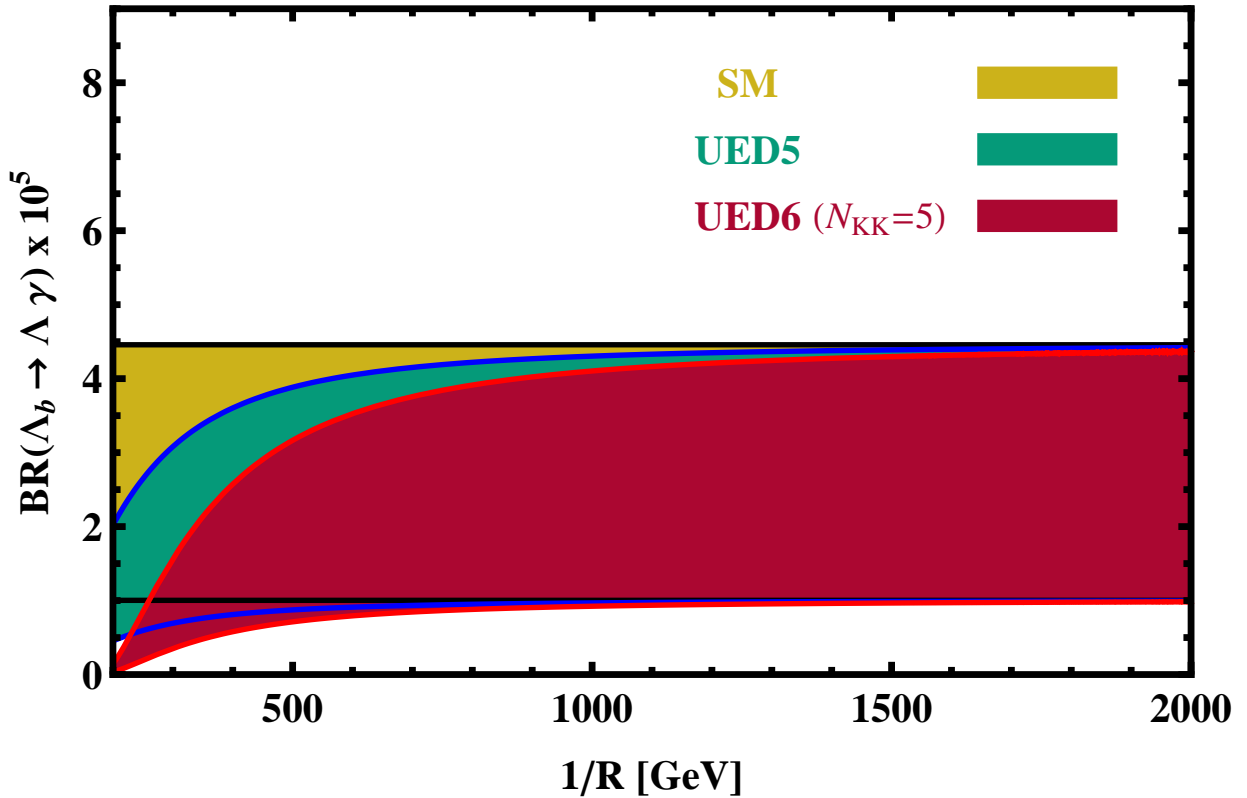,width=1.0\linewidth,clip=}
\end{tabular}
\caption{The dependence of branching ratio on compactification factor $1/R$ for $\Lambda_b
\rightarrow \Lambda \gamma$ decay in SM, UED5 and UED6 with $N_{KK}=5$  when the uncertainties of the form factors are considered.}
\end{figure}
\begin{figure}[h!]
\centering
\begin{tabular}{ccc}
\epsfig{file=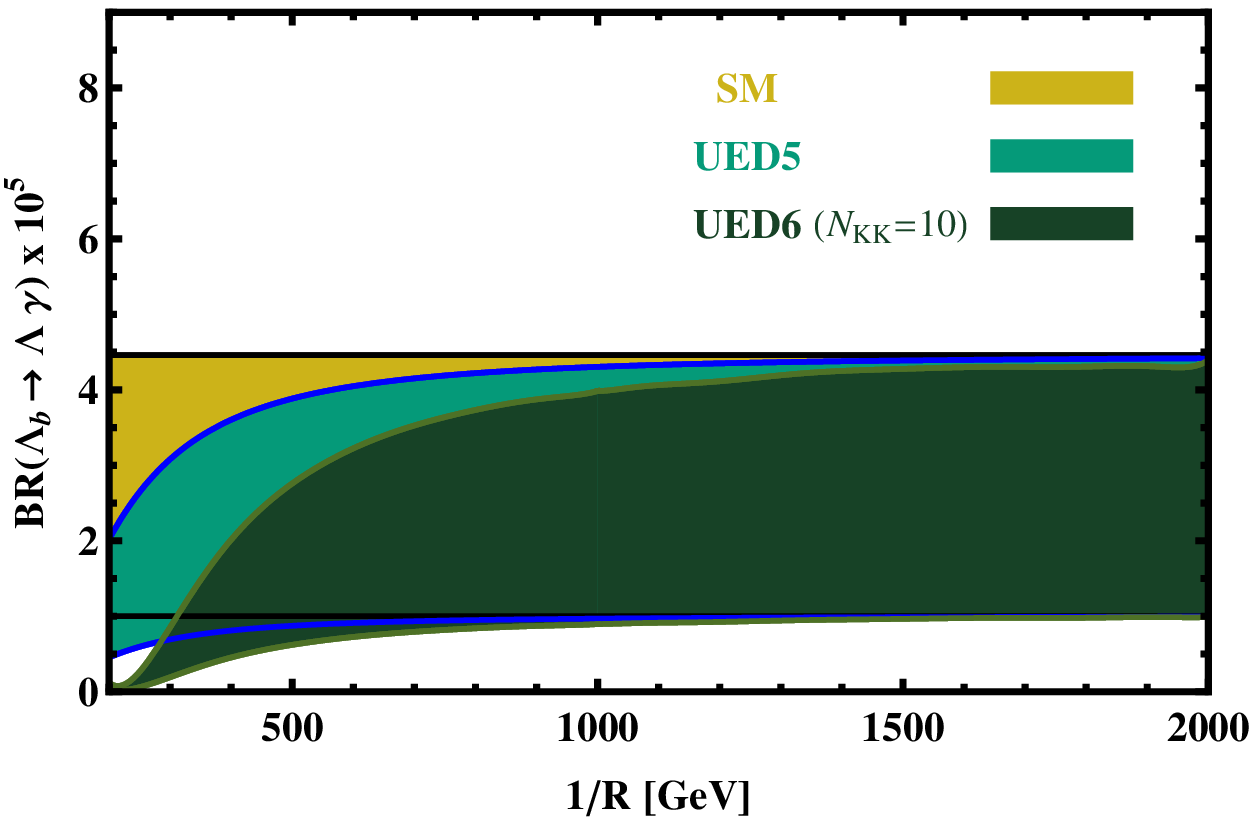,width=1.0\linewidth,clip=}
\end{tabular}
\caption{The same as figure 2 but for $N_{KK}=10$.}
\end{figure}
\begin{figure}[h!]
\centering
\begin{tabular}{ccc}
\epsfig{file=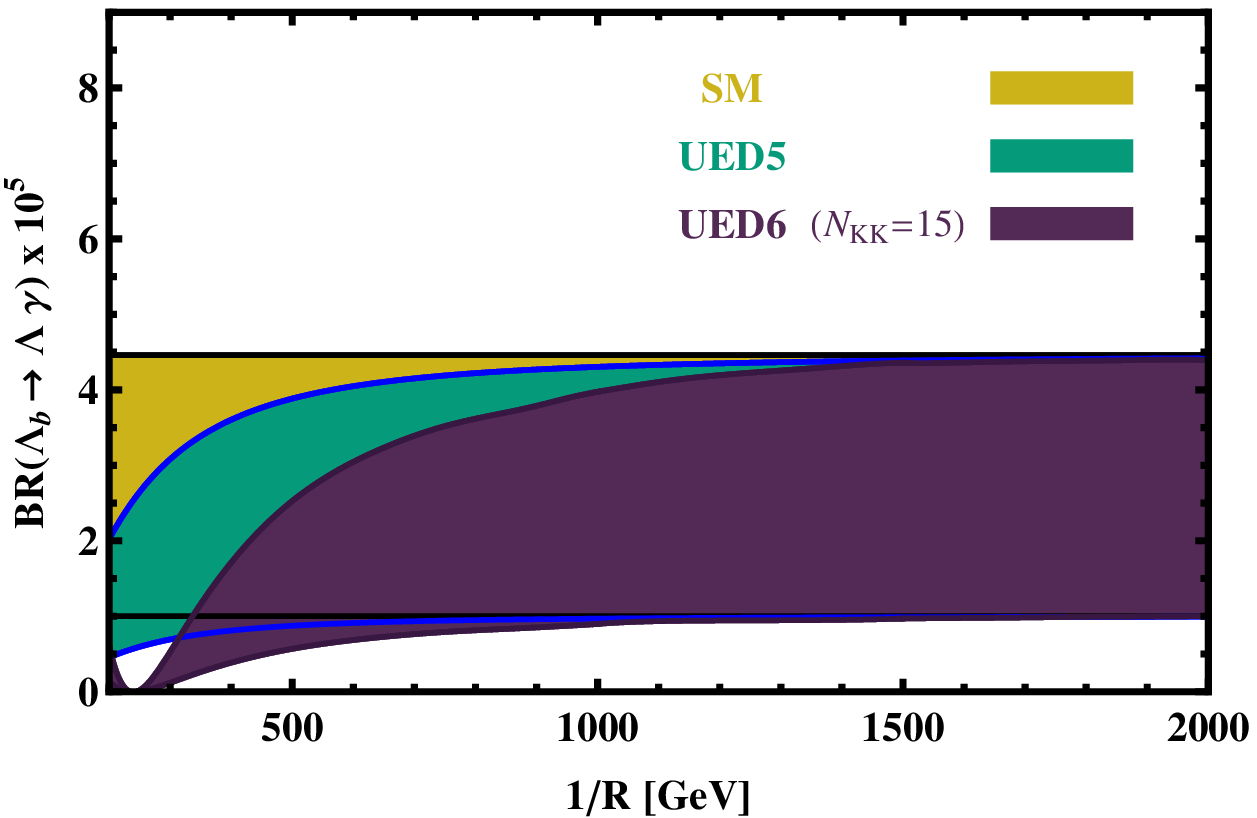,width=1.0\linewidth,clip=}
\end{tabular}
\caption{The same as figure 2 but for $N_{KK}=15$.}
\end{figure}
%
%

At the end of this section, we present the dependence of the branching ratio on $1/R$ considering the errors of  form factors in figures 2, 3 and 4. From these figures we read that the errors of form
 factors can not totally kill the differences between the predictions of the UED models on the branching ratio of $\Lambda_b \rightarrow \Lambda \gamma$ channel
 with that of the SM  at lower values of the compactification scale. These discrepancies can also be seen in the
lower limits favored by  different FCNC transitions in UED5 and UED6 models, cosmological constraints, electroweak precision tests \cite{Biancofiore-Fazio,Cosm-Belanger,Haisch,Gogoladze} as well as
 the latest results of the Higgs search/discovery at the LHC and of the electroweak precision data for the
S and T parameters \cite{takuya} for UED5. However, when $1/R$ approaches to $1~TeV$
all differences of the UED results with the SM predictions are roughly  killed and there are no considerable deviations of the UED predictions from that of the SM at $1.41~TeV$ 
 quoted via direct searches at ATLAS Collaboration \cite{ATLAS} for the $\Lambda_b \rightarrow \Lambda \gamma$ decay channel.
\section{Conclusion}
In the present work, we have performed a comprehensive analysis of the $\Lambda_b \rightarrow \Lambda \gamma$ decay channel in the SM, UED5 and UED6 scenarios. 
In particular, we calculated the total decay rate and  branching ratio for this channel in different UED scenarios and looked for the deviations of the results from the SM predictions. 
We used the expression of the Wilson coefficient $C^{eff}_{7}$ entered to the low energy effective Hamiltonian calculated in SM, UED5 and UED6 models. We also used the numerical values of the form factors 
calculated via light cone QCD sum rules in full theory as the main inputs of the numerical analysis. We detected  considerable discrepancies between the considered UED models' 
predictions with that of the SM prediction at lower values of the compactification factor. These discrepancies can not totally be killed by the uncertainties of the form factors at lower values of $1/R$ and they exist
 at the lower limits favored by  different FCNC transitions in UED5 and UED6 models, cosmological constraints, electroweak precision tests \cite{Biancofiore-Fazio,Cosm-Belanger,Haisch,Gogoladze} as well as
  the latest results of the Higgs search/discovery at the LHC and of the electroweak precision data for the
S and T parameters \cite{takuya}.
However, when $1/R$ approaches to $1~TeV$
all deviations of the UED results from the SM predictions are roughly  killed and there are no considerable deviations of the UED predictions for the $\Lambda_b \rightarrow \Lambda \gamma$ decay channel from that of the SM at $1.41~TeV$ 
 quoted via direct searches at ATLAS Collaboration \cite{ATLAS}.
 The order of branching ratio for  $\Lambda_b \rightarrow \Lambda \gamma$  decay channel in SM shows that this channel 
can be accessible at LHCb. 

\underline{Note Added}: \textit{ After completing this work, a related study  titled as ``Bounds on the compactification scale of two universal extra dimensions
 from exclusive $b \to s \gamma$ decays'' was submitted to arXiv on 28 Feb 2013 with arXiv:1302.7240 [hep-ph] \cite{Pietro}, where a 
similar analysis is done only in UED6 using the form factors calculated from the heavy quark effective theory and average value of the $N_{KK}$. When
 we compare our results with those of \cite{Pietro}, we see that there is a considerable difference between our result on the  branching ratio of the decay under consideration in SM with those of \cite{Pietro}. Although the central values of 
  the branching ratios in two works obtained via UED6 have similar behaviors, the bands of UED6 in our case sweep wide ranges compared to those of \cite{Pietro}. 
Especially, the band of UED6 ($N_{KK}=10$) in \cite{Pietro} starts to completely cover the SM band at  $1/R\approx800~GeV$, while
 in our case, we see a similar behavior at $1/R\approx1000~GeV$. These small differences can be attributed to different form factors used in the numerical analysis as well as other input 
parameters. }

\section{Acknowledgement}
 We would like to thank A. Freitas and U. Haisch for useful discussions. 
 
\end{document}